\documentclass[aps,pre,amsmath,amsfonts,amssymb,twocolumn]{revtex4}
\usepackage{bm}        

\usepackage[latin1]{inputenc}
\usepackage{graphicx}
\usepackage[english]{babel}
\usepackage[usenames,dvipsnames]{color}

\usepackage{amsfonts}
\usepackage{amsmath}
\usepackage{amssymb}

\usepackage{color}

\usepackage{soul}

\newcommand{\beq}{\begin{equation}}
\newcommand{\eeq}{\end{equation}}
\newcommand{\nn}{\nonumber}
\newcommand{\ket}[1]{|#1\rangle}
\newcommand{\bra}[1]{\langle #1|}

\newcommand{\ra}{\rightarrow}

\makeatletter
\@ifundefined{textcolor}{}
{%
 \definecolor{BLACK}{gray}{0}
 \definecolor{WHITE}{gray}{1}
 \definecolor{RED}{rgb}{1,0,0}
 \definecolor{GREEN}{rgb}{0,1,0}
 \definecolor{BLUE}{rgb}{0,0,1}
 \definecolor{CYAN}{cmyk}{1,0,0,0}
 \definecolor{MAGENTA}{cmyk}{0,1,0,0}
 \definecolor{YELLOW}{cmyk}{0,0,1,0}
 }

\begin{document}


\title{Emergence of energy-avoiding and energy-seeking behaviours in \\ nonequilibrium dissipative quantum systems}

\author{Thiago Werlang
$^{1}$
}

\author{Maur\'icio Matos
$^{1}$
}

\author{Frederico Brito
$^{2}$
}

\author{Daniel Valente
$^{1}$
}
\email{valente.daniel@gmail.com}

\affiliation{
$^{1}$ 
Instituto de F\'isica, Universidade Federal de Mato Grosso, Cuiab\'a, 78060-900, Mato Grosso, Brazil
}

\affiliation{
$^{2}$ 
Instituto de F\'isica de S\~ao Carlos, Universidade de S\~ao Paulo, Caixa Postal 369, S\~ao Carlos, 13560-970, S\~ao Paulo, Brazil
}

\maketitle
\section{Abstract}
A longstanding challenge in nonequilibrium thermodynamics is to predict the emergence of self-organized behaviours and functionalities typical of living matter.
Despite the progress with classical complex systems, it remains far from obvious how to extrapolate these results down to the quantum scale.
Here, we employ the paradigmatic master equation framework to establish that some lifelike behaviours and functionalities can indeed emerge in elementary dissipative quantum systems driven out of equilibrium.
Specifically, we find both energy-avoiding (low steady dissipation) and energy-seeking behaviours (high steady dissipation), as well as self-adaptive shifts between these modes, in generic few-level systems.
We also find emergent functionalities, namely, a self-organized thermal gradient in the system's environment (in the energy-seeking mode) and an active equilibration against thermal gradients (in the energy-avoiding mode).
Finally, we discuss the possibility that our results could be related to the concept of dissipative adaptation.

\section{Introduction}
When matter is driven far from thermal equilibrium and dynamically dissipates energy, self-organized states and processes may emerge.
Examples are as diverse as tornadoes and living cells.
The kind of self-organization that has led to the transition from inanimate to living matter may look distinctive perhaps in part due to the emergence of seemingly purposeful functionalities and finely-tuned adaptive behaviours \cite{petra17,nnano2020}.
We are thus left with the challenge of looking for general principles behind the emergence of exceptional behaviours that look like those found in living organisms \cite{nnano2015}.
From a fundamental viewpoint, this can be significant for a better understanding of life as we know it.
For instance, because the notion of natural selection does not apply to situations prior to the existence of the first living cells, one may turn to nonequilibrium statistical mechanics \cite{goldenfeld06} and thermodynamics \cite{prx16} as attempts to generalize Darwinian evolution.
From a practical viewpoint, searching for physical principles may allow us to extend to diverse systems the self-organized functionalities typically found in living matter.
Studies with classical many-body systems have recently made progress in this direction \cite{nphys12, PRE2015,nnano2015,huck16,ncomm17,PRL2017,PNAS2017,nmat2017,ragazzon18,nphoton2018,kedia19,nnano2020,nphys2020}, showing, in particular, the emergence of energy-seeking \cite{PRE2015,PRL2017} and of energy-avoiding behaviours \cite{PRL2017,nmat2017,kedia19}, of dynamical self-healing \cite{PRE2015,ncomm17,nmat2017}, and of self-adaptation under certain changes in the environment \cite{nphys12,nmat2017,ncomm17,nphoton2018}.
While, on the one hand, we may expect emergent lifelike properties to require some degree of complexity and many-body systems, on the other hand, in our search for principles we usually try to simplify things as far as possible, keeping the essential ingredients only.
Finding just the right balance is all but a trivial task: quoting Schwille, how simple could life be? \cite{petra17}.

Here, we look for the simplest possible scenarios where those lifelike self-organized behaviours and functions can emerge, with special interest in extending them to the quantum scale.
We consider the dissipative dynamics of generic driven quantum systems, as described by the well-established Markovian quantum master equation.
We show self-organized energy-avoiding behaviour (leading to low dissipation of energy in the steady state) and energy-seeking behaviour (leading to high dissipation) in three-level systems.
We also show a self-adaptation between these two modes in a four-level system, depending on a change of stimulus.
We find that, when the four-level system is in its energy-seeking mode, it can build a self-organized thermal gradient in its own environment.
In its energy-avoiding mode, by contrast, the four-level system is shown to act against thermal gradients, actively seeking to equilibrate its own environment.
Finally, we sketch a plausible link between our results and the notion of dissipative history, as conveyed by the concept of  dissipative adaptation \cite{nnano2015,nnano2020,cp21}.
Throughout the text, we present biologically-inspired processes in order to support our hypothesis that some lifelike behaviours can emerge in far from equilibrium matter even in the absence of natural selection.

\section{Results}

\subsection{Energy-avoiding behaviour}
Living organisms have evolved to keep themselves structurally stable and functional.
To maintain stability, one way used by living systems is to avoid absorbing energy of specific types.
Bones, shells and cell walls are specially adapted to guarantee mechanical stability against wounds, just as photosynthetic pigments are well-suited to avoid the damaging absorption of ultraviolet light.
From a broad thermodynamic perspective \cite{nnano2015,nnano2020}, energy-avoiding structures may self-assemble when a physical system finds stable states achieved by means of a history of nonequilibrium work absorption and dissipation.
These stable states have, in turn, very low susceptibility to further absorption from that specific nonequilibrium source.
The system then becomes finely tuned to avoid absorbing more of that specific nonequilibrium energy input that drove the transition in the first place \cite{PRL2017,nmat2017,kedia19}.
This is the energy-avoiding mechanism, which can also be deemed a negative feedback loop \cite{nphys12,huck16}.

\begin{figure}[!htb]
\centering
\includegraphics[width=1.0\linewidth]{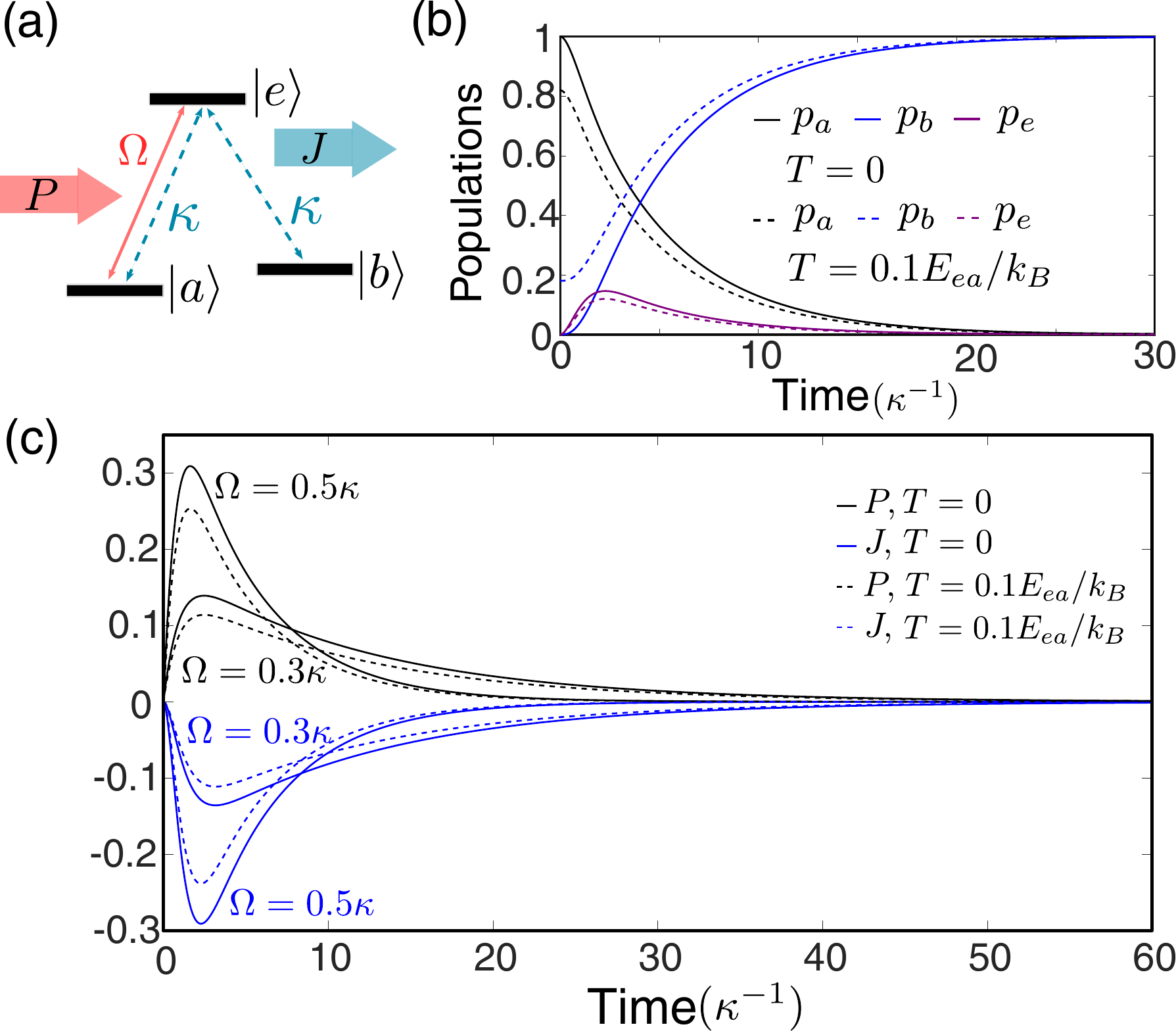} 
\caption{
{\bf Energy-avoiding behaviour in a three-level system.}
(a) $\Lambda$ system, with eigenstates $\ket{e}$, $\ket{b}$ and $\ket{a}$ (black horizontal bars), with corresponding eigenenergies $E_e \gg E_b > E_a$.
The spontaneous emission rate of each transition is $\kappa$ (dashed arrows), and the resonant drive with the coupling strength $\Omega$ induces transitions between states $\ket{a}$ and $\ket{e}$ (full arrow).
Energy exchange rates are characterized as power consumption $P$ and dissipation $J$.
(b) Departing from thermal equilibrium (at times $t < 0$), the pump is suddenly turned on and kept constant (at $t\geq 0$), with $\Omega = 0.5 \kappa$, so the state of the system asymptotically self-organizes towards a nonequilibrium state, with populations 
$p_b(\infty) \ra 1 \gg p_b(0)$ and 
$p_a(\infty) \ra 0 \ll p_a(0)$. 
In the meantime, the system undergoes a finite excitation probability, $p_e(t) > 0$.
(c) In the transient regime, both the absorption $P(t)$ (black) and the dissipation $J(t)$ (blue) are appreciable (both in units of $E_{ea}\kappa$, where $E_{ea} = E_e - E_a$).
Asymptotically, the stationary self-organized nonequilibrium state lowers the power consumption and the absolute value of the dissipation, $P(\infty) = |J(\infty)| \ra 0$.
In (b) and (c), we compare $T = 0$ (solid curves) with the finite (low) temperature $T = 0.1 E_{ea} /k_\mathrm{B}$ (dashed curves).
}
\label{fig1}
\end{figure}

Here, we look for the emergence of this type of energy-avoiding mechanism, as described above, in the dynamics of an elementary physical system.
Specifically, we search for an open system that departs from thermal equilibrium and, once driven by a nonequilibrium source, transiently absorbs enough energy so as to finally stabilize in a nonequilibrium steady state characterized by a strong reduction in the amount of energy steadily absorbed and dissipated from that given drive.
We find this energy-avoiding behaviour emerging in a lambda ($\Lambda$) three-level system.
Let us label the energy eigenstates as $\ket{a}$, $\ket{b}$ and $\ket{e}$, in increasing order, as illustrated in Fig.(\ref{fig1})(a).
We suppose that, initially (at times $t<0$), the $\Lambda$ system is in thermal equilibrium with an environment at temperature $T$.
A nonequilibrium energy source, more specifically, a classical resonant monochromatic light beam, is suddenly turned on and kept constant (at $t \geq 0$).
As a result, detailed balance is broken, and the system undergoes a dissipative dynamics towards a self-organized nonequilibrium quantum state.
A self-organized state here means an atypical (exceptional) stationary quantum state (as compared with the thermal equilibrium state of the same system), achieved autonomously (i.e, with no pre-determined goal encoded in the energy source).
To emphasize the energetic exchanges driving the dynamics, we recast the master equation as
\beq
\partial_t p_{e} = P/E_{ea} + \sum_{m} J_{em}/E_{em},
\eeq
where $P$ is the absorbed power (at the origin of broken detailed balance), $E_{em} = E_e-E_m$ is the energy gap between the states $\ket{e}$ and $\ket{m} = \ket{a}, \ket{b}$, $J_{em}$ is the dissipation current in the transition from $\ket{e}$ to $\ket{m}$, and $p_e(t) \equiv \rho_{ee}(t)$ is the excitation probability (with $\rho_{ij} \equiv \bra{i} \rho \ket{j}$, and $\rho$ being the density operator of the system).
In turn, the state of this nonlinear, bistable system strongly affects back how energy will be absorbed.
That is, the power absorption,
\beq
P = 2 E_{ea} \Omega \ \mathrm{Im}(\rho_{ae}),
\label{P}
\eeq
depends on the state of the system, specially on the imaginary part of the quantum coherence $\rho_{ae}$ between the states $\ket{a}$ and $\ket{e}$, whereas the dissipation rates,
\beq
J_{em} = - E_{em} \kappa [(1+n_{em}) \rho_{ee} - n_{em} \rho_{mm}],
\label{Jem}
\eeq
arise from stochastic transitions, involving the populations of all states, namely, $\ket{a}$, $\ket{b}$, and $\ket{e}$.
In Eq.(\ref{P}), $\Omega$ is the system-field coupling strength (aka the Rabi frequency, proportional to the amplitude of the incoming field).
In Eq.(\ref{Jem}), $\kappa$ is the spontaneous emission rate, and $n_{em}$ describes a Bose-Einstein distribution.
This closes a nonlinear (state-dependent) feedback loop, where energy consumption leads to a dynamical change, which strongly affects back how much energy is absorbed.
At all times, $J_{em}$ provides a measure of broken detailed balance (see further details in the Methods).

After the transient absorption peak, the system achieves the (nonequilibrium) stationary state
\beq
p_b(\infty) \approx 1,
\label{selforg}
\eeq
with equality $p_b(\infty) = 1$ holding at $T = 0$, where $p_b(t) \equiv \rho_{bb}(t)$ is the population of state $\ket{b}$.
We call this a self-organized state, since we are considering $E_b > E_a$ (which means that Eq.(\ref{selforg}) is quite far from the thermal-equilibrium values for the populations, namely, $p_b \rightarrow 0$ and $p_a \rightarrow 1$, in the $T \rightarrow 0$ limit).
See Fig.(\ref{fig1})(b).
In the stationary regime, the self-organized nonequilibrium state makes both $P$ and the total dissipation $|J|$, where $J \equiv \sum_m J_{em}$, to converge towards very low values, as we see in Fig.(\ref{fig1})(c).
In other words, the system seems specially adapted to avoid absorbing (hence to avoid dissipating) more of that incoming light that drove the transition in the first place.
We note that, at precisely $T = 0$, detailed balance is reestablished in the steady-state, $P(\infty) = J_{ea}(\infty) = J_{eb}(\infty) = 0$, although a small power consumption, $P(\infty) = - J(\infty) \approx E_{ea} \kappa n_{eb}$, still holds at very low temperatures ($n_{eb} \ra 0$).
Fig.(\ref{fig1})(c) also shows that higher coupling strengths $\Omega$ accelerate the system's adaptation towards this energy-avoiding behaviour of low steady $P$.
In Fig.(\ref{fig1})(b) and (c), we compare the zero-temperature limit, $T = 0$, with the finite temperature $T = 0.1 E_{ea} /k_\mathrm{B}$, where $k_\mathrm{B}$ is the Boltzmann constant.
This shows that the energy-avoiding behaviour is homeostatically robust against thermal fluctuations, at low-enough temperatures.

\subsection{Energy-seeking behaviour}
Living cells have also developed a sophisticated network of processes, called metabolism, characterized by the capacity of the cell to absorb specific types of nonequilibrium energy available in its environment, and then use it to maintain its (nonequilibrium) living state.
Coming back to our broad thermodynamic perspective, a generic fluctuating physical system driven far from equilibrium may also dynamically (dissipatively) self-adjust towards an exceptional stationary state that seems specially adapted to maintain high rates of energy consumption from the specific drive that led the system to that nonequilibrium state in the first place \cite{PRE2015,PRL2017}.
Such an emergent energy-seeking behaviour can also be regarded as a positive feedback loop \cite{nphys12,huck16}.

\begin{figure}[!htb]
\centering
\includegraphics[width=1.0\linewidth]{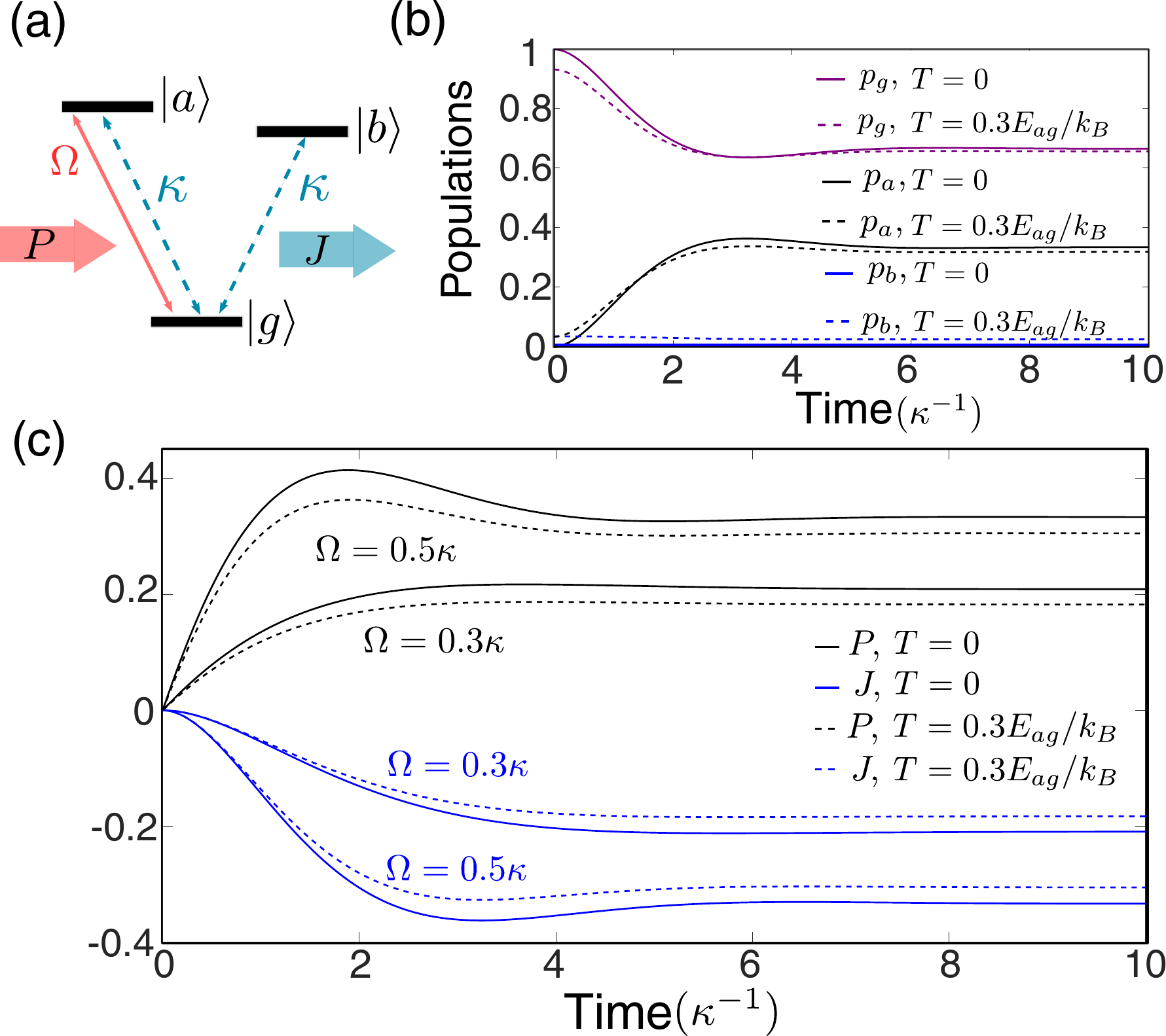} 
\caption{
{\bf Energy-seeking behaviour in a three-level system.}
(a) $V$ system with eigenstates $\ket{a}$, $\ket{b}$ and $\ket{g}$ (black horizontal bars), and corresponding eigenenergies $E_a > E_b \gg E_g$.
The spontaneous emission rate of each transition is $\kappa$ (dashed arrows) and the resonant drive induces the coupling between states $\ket{a}$ and $\ket{g}$ with strength $\Omega$ (full arrow).
Energy exchange rates are characterized as power consumption $P$ and dissipation $J$.
(b) Departing from thermal equilibrium (at times $t<0$), the pump is turned on and kept constant (at $t\geq 0$), with $\Omega = 0.5 \kappa$, so the state of the system asymptotically self-organizes towards a nonequilibrium state where the populations satisfy $p_a(\infty) > p_a(0)$ and $p_g(\infty) < p_g (0)$.
(c) After a transient regime, the stationary self-organized state maintains high power consumption and dissipation, $P(\infty) = |J(\infty)| > 0$ (both in units of $E_{ag}\kappa$, where $E_{ag} = E_a - E_g$), which increase with $\Omega$.
In (b) and (c), we compare $T = 0$ (solid curves) with the finite (low) temperature $T = 0.3 E_{ag} /k_\mathrm{B}$ (dashed curves).
}
\label{fig2}
\end{figure}

We find this energy-seeking behaviour in a three-level quantum system, but in $V$ configuration.
We label the energy eigenstates as $\ket{a}$, $\ket{b}$, and $\ket{g}$, in decreasing order (see Fig.(\ref{fig2})(a)).
Let us consider that, at times $t<0$, the $V$ system is at thermal equilibrium.
By pumping one of its transitions with a classical resonant monochromatic light at $t \geq 0$, the $V$ system undergoes a transition to a self-organized state, which seems to be exceptionally adapted to keep absorbing more energy from the nonequilibrium source.
That is, the population $p_a(\infty) \equiv \rho_{aa}(\infty)$ increases with respect to its equilibrium value (see Fig.\ref{fig2}(b)), and a quantum coherence between the states $\ket{a}$ and $\ket{g}$, namely $\rho_{ag}(\infty)$, is built up (see Methods), as if the system was purposefully trying to keep high stationary power consumption.
In turn, the system dissipates high amounts of energy in a steady manner, maintaining its nonequilibrium state.
Figure (\ref{fig2})(c) shows the dynamics of the absorption power $P$ and the dissipation $J$ as functions of time, at distinct coupling strengths $\Omega$.
The higher the $\Omega$, the more power is steadily absorbed from light and the higher is the dissipation.
Similarly to the energy-avoiding mode, the energy-seeking behaviour here is reasonably resilient against thermal perturbations, as shown in the comparison between $T = 0$ and $T = 0.3 E_{ag}/k_\mathrm{B}$.

\subsection{Self-adaptative shifts}
\begin{figure}[!htb]
\centering
\includegraphics[width=1.0\linewidth]{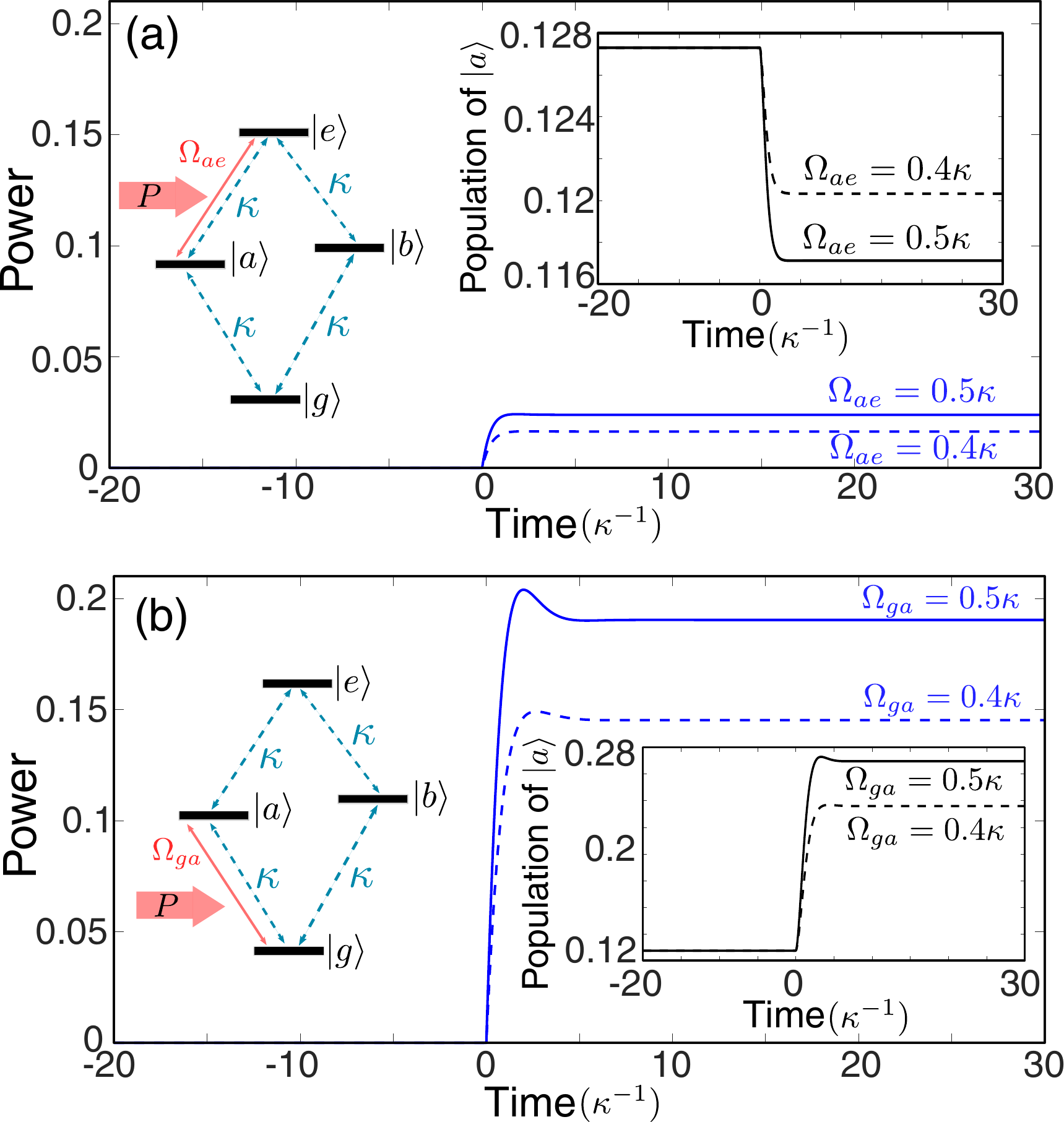} 
\caption{
{\bf Self-adaptive shift between energy-avoiding and energy-seeking behaviours in a four-level system.}
(a) $\Diamond$ system with eigenstates $\ket{g}$, $\ket{a}$, $\ket{b}$ and $\ket{e}$ (black horizontal bars), and corresponding eigenenergies $E_g \ll E_a < E_b \ll E_e$.
The spontaneous emission rate of each transition is $\kappa$ (dashed arrows) and the resonant drive induces the coupling between states $\ket{a}$ and $\ket{e}$ with strength $\Omega_{ae}$ (full arrow).
$P$ stands for the power consumption (units of $E_{ea}\kappa$, where $E_{ea} = E_e - E_a$).
At times $t<0$, the diamond system is at thermal equilibrium.
At $t\geq 0$ the $\Lambda$-type transition undergoes stimulation ($\Omega_{ae} > 0$), making the energy-avoiding behaviour to emerge, so that the stationary power absorption $P$ is low and almost insensitive to the increase from $\Omega_{ae} = 0.4 \kappa$ (dashed blue) to $0.5 \kappa$ (solid blue).
(Inset) the stationary population of state $\ket{a}$ decreases, as if the system was intentionally avoiding absorbing higher powers.
(b) Same $\Diamond$ system as in (a), but with the resonant drive coupling states $\ket{g}$ and $\ket{a}$ with strength $\Omega_{ga}$ (full arrow).
At times $t<0$, the diamond system is at thermal equilibrium.
At $t\geq 0$ the $V$-type transition undergoes stimulation ($\Omega_{ga} > 0$), making the energy-seeking behaviour to emerge, so that the stationary power absorption $P$ is high and quite sensitive to the increase from $\Omega_{ae} = 0.4 \kappa$ (dashed blue) to $0.5 \kappa$ (solid blue).
(Inset) the stationary population of state $\ket{a}$ increases, as if the system was intentionally seeking to absorb higher powers.
In all plots, we consider the temperature $T = 0.5 E_{ea} / k_\mathrm{B}$.
}
\label{fig3}
\end{figure}

Can a few-level quantum system present a versatile behaviour where it autonomously shifts from the energy-seeking to the energy-avoiding mode and back?
Self-adaptive shifts in a system's behaviour \cite{nmat2017,ncomm17,nphoton2018} represent a close analogy with the highly selective, switch-like response of receptors in living organisms \cite{nphys12}.
We implement such a self-adaptive shifting behaviour in a four-level system with a diamond energy structure ($\Diamond$).
Our motivation is that the $\Diamond$ system integrates, in some sense, features from the $\Lambda$ and the $V$ energy-level structures.
We set out to analyze the change in the system's response under two distinct stimuli.
To that end, we compare the case where the power source drives the transition between states $\ket{a}$ and $\ket{e}$ ($\Lambda$-type) with the case where it drives the transition between states $\ket{g}$ and $\ket{a}$ ($V$-type). 
Here, we have labeled the energy levels in increasing order as $\ket{g}$, $\ket{a}$, $\ket{b}$ and $\ket{e}$ (see Fig.(\ref{fig3})).

We find that, as expected, when the $\Lambda$-type transition is driven, the energy-avoiding behaviour emerges (see Fig.(\ref{fig3}), panel (a)), whereas driving the $V$-type transition leads to the energy-seeking behaviour (as in Fig.(\ref{fig3}), panel (b)).
A remark must be made.
In contrast to the results shown in Fig.(\ref{fig1}), we see directly from the blue curves in Fig.(\ref{fig3})(a) that the absorbed power stabilizes at finite values, instead of dropping to (or close to) zero.
This can be seen in Fig.(\ref{fig3})(a), where we plot $P$ as a function of time.
The reason is the finite lifetime of state $\ket{b}$ in the $\Diamond$ configuration. 
By decaying towards $\ket{g}$, the system opens a pathway for a steady non-vanishing dissipation to the environment.
To better clarify that this can indeed be characterized as an energy-avoiding behaviour, we show, in addition to the very low values of power absorption, that the population of state $\ket{a}$ decreases with respect to its thermal equilibrium value.
As we can see from the inset of panel (a) in Fig.(\ref{fig3}), stronger drives tend to empty state $\ket{a}$ more strongly, meaning that the system behaves as if it was purposefully avoiding to absorb energy from the driving light.
Also, we see that the increase in $\Omega_{ae}$ leads to a slight increase in $P$.
In the energy-seeking mode, Fig.(\ref{fig3})(b), we find the opposite behaviour, namely, a stationary increase in the population of state $\ket{a}$ with respect to the equilibrium value.
This is shown in the inset of panel (b).
Finally, panel (b) evidences that not only $P$ is much higher in the energy-seeking mode, but also that the increase in 
$\Omega_{ga}$ leads to a visible increase in the absorption power.

\subsection{Self-organized thermal gradients and active thermalization}
Temperature gradients can be crucial to nonequilibrium self-organization.
For instance, they can play a significant role in the emergence of complex chemistry \cite{gt12,gt15,ncomm17,gt17,gt20,gt21}, with possible implications to the problem of the origin of life on earth \cite{gt17,gt20,gt21}.
Temperature gradients may also result in broken detailed balance, thus competing with the effects of other nonequilibrium sources (the driving light, in our case) \cite{nphoton2018}.
This competition makes it more challenging to achieve homeostatic resilience, as compared to thermal-equilibrium fluctuations.
This alone could be reason enough for us to report on the effects of thermal gradients in this paper.
To our surprise, however, we have found that the emergent behaviours analyzed above imply self-organized functionalities related to thermal gradients: a four-level system can autonomously act upon the temperature gradients in its own environment and modify it.
This reminds us of the apparently purposeful, end-directed actions taken by living organisms when altering their own surroundings, as well as of the measurable thermal gradients within the boundaries of single living cells \cite{gtsc}.

Specifically, we find that the energy-seeking mode results in a self-organized thermal gradient in the environment, as explained below.
Let us suppose that the $\Diamond$ system is coupled to two distinct environments, namely, left ($L$) and right ($R$), both initially at thermal equilibrium.
Environment $L$ is the one coupled to the $\ket{g}$-$\ket{a}$-$\ket{e}$ transitions, whereas $R$ is coupled to the $\ket{g}$-$\ket{b}$-$\ket{e}$ transitions.
The above assumption is inspired by models of energy transport through small quantum chains and their applications in biophysical scenarios \cite{w14,qc2,qc3,qc4}.
When the ($V$-type) transition $\ket{g}$-$\ket{a}$ is stimulated, we find that the energy-seeking mode emerges.
The system thus preferentially dissipates towards $L$, that is, $|J_L| > |J_R|$ 
(here, $J_L \equiv J_{ea} + J_{ag}$ and $J_R \equiv J_{eb} + J_{bg}$).
The key point is that, if the heat capacities of both $L$ and $R$ are finite and comparable, the left temperature will raise faster, implying that $T_L > T_R$ at some point.
Once $L$ is warmer, the system will keep favoring dissipation towards $L$, therefore tending to increase the disequilibrium even further.
We also find that the system preferentially dissipates towards environment $L$ even if the temperature gradient is initially (externally) set against the natural tendency of the self-organized gradient, that is, when $T_L$ is initially lower than $T_R$.

\begin{figure}[!htb]
\centering
\includegraphics[width=1.0\linewidth]{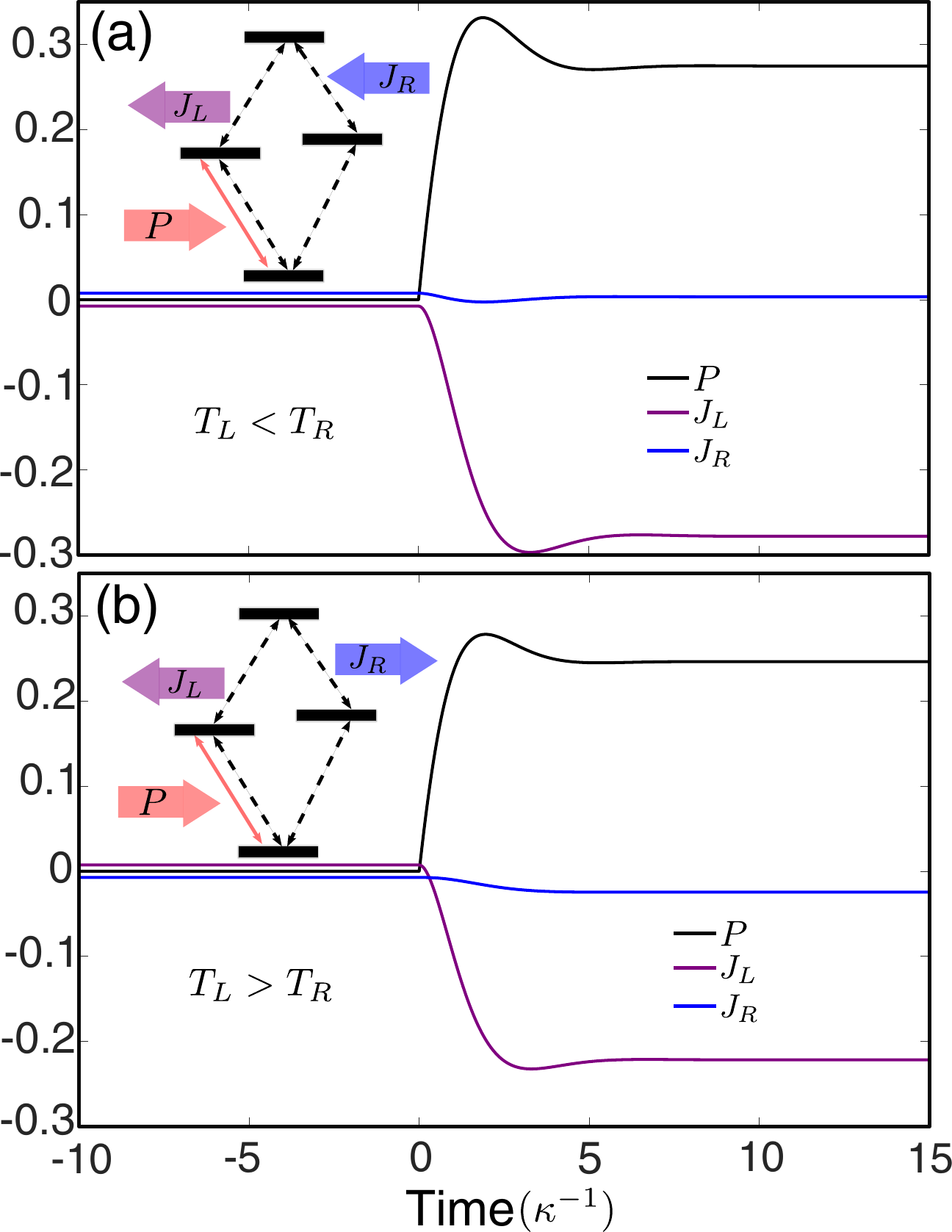} 
\caption{
{\bf Self-organized thermal gradient in the energy-seeking mode.}
(a) $\Diamond$ system with eigenstates $\ket{g}$, $\ket{a}$, $\ket{b}$ and $\ket{e}$ (black horizontal bars), and eigenenergies $E_g \ll E_a < E_b \ll E_e$.
Spontaneous emission transitions (dashed arrows) are considered with rates $\kappa$. 
The coupling strength with the resonant drive (full arrow) is set to $\Omega_{ga} = 0.5 \kappa$ (at times $t\geq 0$).
Power consumption is characterized by $P$.
Heat currents are characterized by $J_L$ and $J_R$, such that $J_{L(R)}$ is coming from the left (right) environment, at temperature $T_{L(R)}$.
We plot $P(t)$ (black), $J_L(t)$ (purple), and $J_R(t)$ (blue), in units of $E_{ea} \kappa$ (where $E_{ea} = E_e - E_a$).
We also set $T_L = 0.2 E_{ea} / k_\mathrm{B}$ and $T_R = 0.4 E_{ea} / k_\mathrm{B}$.
(b) Same as in (a), but with $T_L = 0.4 E_{ea} / k_\mathrm{B}$ and $T_R = 0.2 E_{ea} / k_\mathrm{B}$.
In both (a) and (b), dissipation is higher towards environment $L$, i.e., $|J_L(\infty)| > |J_R(\infty)|$, thus enabling a self-organized thermal gradient (towards $T_L > T_R$), and consuming high powers.
}
\label{fig4}
\end{figure}

Figures (\ref{fig4}) (a) and (b) characterize this preferential dissipation towards environment $L$.
In (a), we set $T_L < T_R$.
At times $t<0$, the stimulating field is off (so that $P=0$), and the heat currents follow standard steady thermal conduction ($J_R = -J_L > 0$), thermally breaking detailed balance.
At $t \geq 0$, the stimulating field is turned on and kept constant, so that high power consumption is achieved.
Because this power consumption leads to increased jumps between states $\ket{a}$ and $\ket{g}$, the dissipation $J_{ag}$ becomes predominant, hence $|J_L|>|J_R|$.
In (b), we reverse the thermal gradient, setting $T_L > T_R$.
At times $t<0$, the absence of the light field makes for $J_L = - J_R > 0$, as expected.
At $t \geq 0$, the high power consumption still leads to higher $J_{ag}$.
Therefore the net dissipation is again more favorable towards $L$, i.e., $|J_L| > |J_R|$.

We find the opposite behaviour when the ($\Lambda$-type) transition $\ket{a}$-$\ket{e}$ is stimulated. 
In that case, the energy-avoiding mode emerges.
If the environments have initially equal temperatures, the dissipation towards $R$ will be slightly higher (not shown).
Higher dissipation towards $R$ leads to an increase in $T_R$.
However, as soon as $T_R$ becomes sufficiently higher than $T_L$, dissipation towards environment $L$ turns out becoming higher than that towards $R$ (as further discussed in the following paragraph).
That is, the dissipation rates become reversed, $|J_L| > |J_R|$ (in comparison to the $T_L = T_R$ case, in which $|J_L| < |J_R|$).
Due to this higher dissipation towards $L$, the temperature $T_L$ is now led to increase faster.
As a consequence, this will eventually lead to $T_L > T_R$.
When $T_L > T_R$, the dissipation rates will readjust themselves again, turning back to $|J_R| > |J_L|$ (as also discussed in the following paragraph).
To sum up, we find that, whenever there is a thermal gradient, dissipation will always be favored towards the coldest environment, in the energy-avoiding mode.
In other words, the system actively seeks to reestablish thermal equilibrium in its own environment.

\begin{figure}[!htb]
\centering
\includegraphics[width=1.0\linewidth]{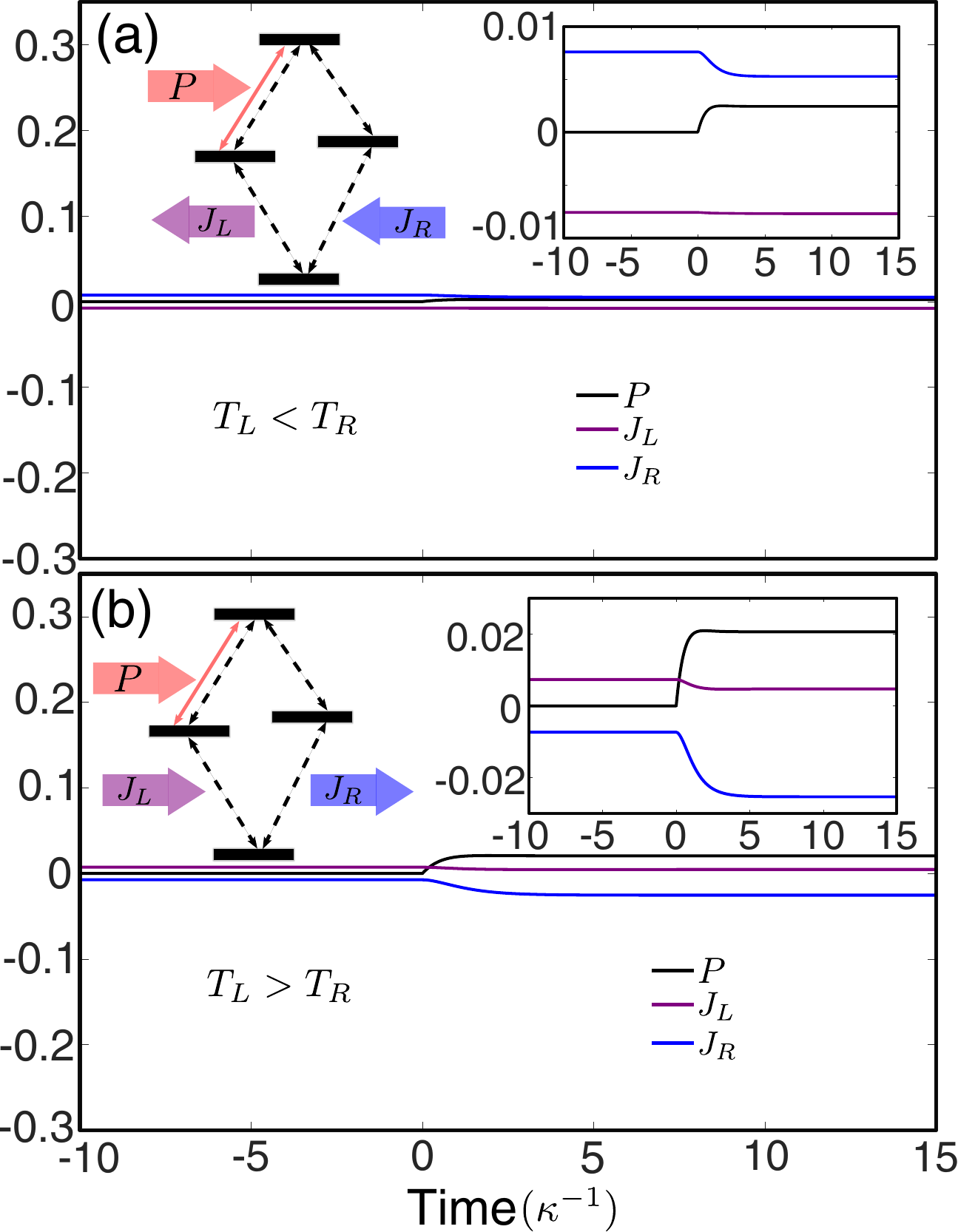} 
\caption{
{\bf Active thermalization in the energy-avoiding mode.}
(a) $\Diamond$ system with eigenstates $\ket{g}$, $\ket{a}$, $\ket{b}$ and $\ket{e}$ (black horizontal bars), and eigenenergies $E_g \ll E_a < E_b \ll E_e$.
Spontaneous emission transitions (dashed arrows) are considered with rates $\kappa$. 
The coupling strength with the resonant drive (full arrow) is set to $\Omega_{ae} = 0.5 \kappa$ (at times $t\geq 0$).
Power consumption is characterized by $P$.
Heat currents are characterized by $J_L$ and $J_R$, such that $J_{L(R)}$ is coming from the left (right) environment, at temperature $T_{L(R)}$.
We plot $P(t)$ (black), $J_L(t)$ (purple), and $J_R(t)$ (blue), in units of $E_{ea} \kappa$ (where $E_{ea} = E_e - E_a$).
We also set $T_L = 0.2 E_{ea} / k_\mathrm{B}$ and $T_R = 0.4 E_{ea} / k_\mathrm{B}$.
(b) Same as in (a), but with $T_L = 0.4 E_{ea} / k_\mathrm{B}$ and $T_R = 0.2 E_{ea} / k_\mathrm{B}$.
In both (a) and (b) (see insets), the dissipation at long times is higher towards the coldest environment, thus inhibiting thermal gradients, and consuming low powers.
}
\label{fig5}
\end{figure}

Figures (\ref{fig5}) (a) and (b) characterize this preferential dissipation towards the coldest environment, as stated in the previous paragraph.
In Fig.(\ref{fig5})(a), we set $T_L < T_R$.
At times $t<0$, the stimulating field is off (so that $P=0$), and the heat currents follow standard steady thermal conduction ($J_R = -J_L > 0$).
At $t \geq 0$, the stimulating field is turned on and kept constant, so that low power consumption takes place (see inset).
Because this power consumption ends up by decreasing the population of state $\ket{a}$ and increasing the population of state $\ket{b}$, the emission towards environment $R$ increases, so that the (positive) difference between the absorbed and the emitted heat decreases, resulting in $|J_L| > |J_R|$.
In Fig.(\ref{fig5})(b), we set $T_L > T_R$.
At times $t<0$, the nonequilibrium field is absent, making for $J_L = - J_R > 0$, as expected from standard thermal conduction.
At $t \geq 0$, the energy-avoiding behaviour again leads to low power consumption (see inset).
Once more, the increased population transfer from $\ket{a}$ to $\ket{b}$ increases the emission towards environment $R$, but now resulting in $|J_R| > |J_L|$, since the heat emission towards $R$ was higher than the absorption from $R$ (the colder environment) in the absence of the drive.
In summary, the plots show that, in the energy-avoiding mode, dissipation is always favored towards the coldest environment, be it $L$ or $R$.

\section{Discussion}
We have established emergent lifelike behaviours and functionalities in elementary, generic open quantum systems.
This represents a step further in the direction of looking for general thermodynamic principles underlying far-from-equilibrium self-organization and the onset of lifelike exceptional behaviours, ultimately allowing us to imitate the transition from inanimate to living matter in diverse systems.
The ubiquity of the master equation framework used here suggests that the disclosed effects can, perhaps, be found in a much wider variety of quantum and classical systems undergoing stochastic dynamics, provided that the systems are sufficiently nonlinear and multistable, therefore giving rise to state-dependent feedback loops, and that the nonequilibrium environments enable transient broken detailed balance.
 
We mention the time-dependent nature of broken detailed balance so as to suggest the relation between the dissipative history and the dynamic responses considered here.
By dissipative history we mean the total (time-integrated) amount of energy dissipated by the system during the process departing from thermal equilibrium and reaching a nonequilibrium steady state.
We hypothesize that, in the context of trying to find a thermodynamic principle characterizing the nonequilibrium stationary state of an arbitrary open physical system, the total dissipated heat may be a relevant quantity (as opposed to the stationary dissipation rate, as analyzed by Kondepudi et al. \cite{PRE2015}, for instance).
We base this hypothesis on two main facts.
The first is that the stationary heat dissipation rate here cannot be a candidate for unifying principle.
To see that, we compare the $\Lambda$ and the $V$ systems.
When subject to the same environmental conditions (equal temperatures and equal-intensity resonant drives), the $\Lambda$ and the $V$ systems are clear examples of opposed behaviours in terms of steady power consumption and dissipation.
The second is that the significance of the time-integrated dissipated heat as a promising candidate has already been discussed in the literature \cite{nnano2015,nnano2020,jcp20,cp21}, giving rise in particular to the concept of dissipative adaptation \cite{nnano2015,nnano2020,cp21}.
Mathematically, the dissipative adaptation has been conceived from Crooks' microscopically reversible condition \cite{crooks}, and establishes a fluctuation theorem where the nonequilibrium work absorption and the heat dissipation along dynamical (transient) trajectories of a system driven far from equilibrium provide a general thermodynamic mechanism explaining driven self-organization (from self-assembly, in particular, to biological adaptation, in general).
The dissipative adaptation paradigm has been recently extended to the quantum realm \cite{cp21}, where the zero-temperature divergences of the classical theorem have been solved.
The models analyzed here are closely related to that quantum dissipative adaptation \cite{cp21}, since the same types of self-organized nonequilibrium quantum states have been obtained (specially with the $\Lambda$ system studied in both papers), but here we have bridged from zero to finite temperatures, comprising thermal gradients, have considered continuous classical light sources (instead of the single-photon pulses), and have achieved self-organized behaviours and functionalities (going beyond self-organized quantum states).
The dissipative histories of the driven $\Lambda$ systems at zero temperature, as analyzed both here and in Valente et al. \cite{cp21}, have been crucial for understanding their self-organized quantum states (see Methods for more details).
Additionally, we highlight that Cook and Endres \cite{jcp20} have also put forward the relevance of the transient over the steady-state dissipation to the stability of chemical non-equilibrium systems described by stochastic dynamical equations.

As a perspective, we would like to further investigate how the dissipative history could be a thermodynamic principle of self-organization across quantum and classical regimes.
Two quantum properties have been relevant in our modeling in this paper, namely, the discreteness of states and the quantum coherences underlying the broken detailed balance.
We believe that quantum coherences, along with quantum-coherent broken detailed balance, will be limited, or even extinguished, by additional environmental noise.
However, incoherent nonequilibrium sources may still enable broken detailed balance in systems described by discrete states, even in the presence of dephasing noises.
In those incoherent scenarios (i.e., those where broken detailed balance does not require quantum superpositions), our results would be equivalent to self-organization in classical multistable systems containing a finite number of local minima.
For that, it suffices that each local minimum is modeled as a discrete, coarse-grained state.
A study in this direction could mean an integration of self-organization across quantum and classical open systems, possibly bridging our results to chemical self-organization and thermophoresis \cite{gt20,gt21}.

\section{Methods}
\section*{Master Equation}
The Hamiltonian of the system is $H_S=\sum_i E_i\ket{i}\bra{i}$. 
In the interaction picture with respect to $H_S$, the dynamics of the system under the Born-Markov approximations is governed by the Lindblad master equation for its density operator $\rho$,
\beq\label{meq}
\partial_t \rho = -(i/\hbar)\left[H_L,\rho \right] + \sum_{i>j} \mathcal{L}_{ij}(\rho).
\eeq
$H_L$ describes the interaction Hamiltonian between the system and the nonequilibrium drive.
The Lindblad superoperator describes the thermal processes,
\begin{eqnarray}
\mathcal{L}_{ij}(\rho)&=& 
\frac{\kappa_{ij}}{2}\left(n_{ij}+1\right)\left[2\sigma_{ji}\rho \sigma_{ij} -\rho \sigma_{ii} -\sigma_{ii}\rho \right] \nn \\&+& \frac{\kappa_{ij}}{2}n_{ij}\left[2\sigma_{ij}\rho \sigma_{ji} -\rho \sigma_{jj} -\sigma_{jj}\rho \right],
\label{lindbladian}
\end{eqnarray}
where $\kappa_{ij}$ are the spontaneous emission rates, 
$\sigma_{ij}\equiv\ket{i}\bra{j}$ are the jump operators,
and each thermal reservoir is at temperature $T_{ij}$, with mean number of excitations given by
$n_{ij}=[\exp(E_{ij}/(k_\mathrm{B} T_{ij}))-1]^{-1}$, 
where $E_{ij} \equiv E_i-E_j$.
In the models analyzed in this paper, we consider classical resonant light fields, in the rotating-wave approximation, so that
$H_L = \sum_{i>j} \hbar \Omega_{ji}(\sigma_{ji} + \sigma_{ij})$, where $\Omega_{ji}$ are the system-field coupling strengths stimulating the transition between states $\ket{j}$ and $\ket{i}$ (also known as Rabi frequencies).

\section*{Energy exchanges}
We assume that the average energy of the system is given by $E\equiv\mbox{Tr}\left(H_S \rho\right)$ in the interaction picture.
Energy exchanges are obtained by using the continuity equation (describing energy conservation), namely
$\partial_t E = \mbox{Tr}\left(H_S\partial_t \rho \right)$,
where we have used that $H_S$ is time-independent.
By applying Eq.(\ref{meq}), we define the input power as the unitary contribution \cite{alicki,prb},
\beq
P \equiv \text{Tr} \big(H_S (-i/\hbar) \left[H_L,\rho \right] \big),
\eeq
and the dissipation rates (heat currents) as the non-unitary contribution \cite{alicki,prb,w14},
\beq
J_{ij} \equiv \text{Tr}\big(H_S \mathcal{L}_{ij}(\rho) \big).
\eeq
By construction, they satisfy
$\partial_t E = P + J$,
where $J \equiv \sum_{i>j}J_{ij}$.

\section*{Feedback loops}
We emphasize here how feedback loops play a key role in our results: 
the power supply first drives the nonlinear system towards nonequilibrium steady-states which, in turn, strongly affect back how much power is absorbed from the drive.
This strong dependence of power absorption on the system's state can be regarded as a kind of state-dependent nonlinear susceptibility.
To make this idea more explicit, we recast Eq.(\ref{meq}), more precisely, the populations $\rho_{nn} \equiv \bra{n} \rho \ket{n}$, in terms of power consumption and dissipation rates as
\begin{align}
\partial_t \rho_{nn} = &\sum_{j<n} (P_{nj}/E_{nj}) - \sum_{i>n} (P_{in}/E_{in}) \nn \\
+ &\sum_{j<n} (J_{nj}/E_{nj}) - \sum_{i>n} (J_{in}/E_{in}).
\label{eqm}
\end{align}
In turn, the power consumption and the dissipation rates depend on the state of the system, that is,
\beq
P_{ij} \equiv E_{ij} \Omega_{ji}[i(\rho_{ij}-\rho_{ji})],
\label{pij}
\eeq
so that $P = \sum_{i>j} P_{ij}$,
and
\beq
J_{ij} = - E_{ij} \Gamma_{ij},
\label{jdb}
\eeq
where we have defined
\beq
\Gamma_{ij} \equiv \kappa_{ij}[(1+n_{ij})\rho_{ii} -n_{ij} \rho_{jj}].
\label{gamaij}
\eeq
Note that $\Gamma_{ij}$ consists of a time-dependent quantitative measure of broken detailed balance.

\section*{$\Lambda$ system}
We consider $i = a, b, e$, so that $E_a < E_b \ll E_e$.
The coupling strengths are such that $\Omega_{ae} = \Omega$ and $\Omega_{ab} = \Omega_{be} = 0$.
The spontaneous emission rates are $\kappa_{ea} = \kappa_{eb} = \kappa$, and $\kappa_{ba} = 0$.
Similarly, the temperatures are $T_{ea} = T_{eb} = T$.
At $T = 0$ and $\Omega = 0$, the steady-state is not uniquely defined, hence introducing a kind of multistability.
At $T = 0$ and $\Omega > 0$, the system self-organizes to the (unique) asymptotic state $\rho(\infty) = \ket{b}\bra{b}$, implying that $P = J_{ea} = J_{eb} = 0$ (hence preserving detailed balance in the stationary regime, whereas breaking it in the transient regime).

To show the equivalence between our results here (coherent drive, at $T=0$) and the single-photon drive analyzed in Valente et al. \cite{cp21}, we combine Eqs.(\ref{eqm})-(\ref{gamaij}) and find that
\beq
\rho_{bb}(\infty)\large|_{\rho_{aa}(0)=1} = W/(2E_{ea}),
\eeq
where the total incoming work is $W \equiv \int_0^\infty P(t')dt'$ and the dynamics is calculated with the initial state $\ket{a}$ (as in Valente et al. \cite{cp21}).
This means that the asymptotic transition probability from $\ket{a}$ to $\ket{b}$ is maximized along with the total work $W$ performed by the coherent field on the system.
Similarly, the total dissipated heat $|Q|$ (where $Q \equiv \int_0^\infty J(t')dt'$) is maximized conditional to the same asymptotic transition probability, namely,
$\rho_{bb}(\infty)\large|_{\rho_{aa}(0)=1} = (|Q| + E_a)/(2 E_{ea} - E_{ba})$ (again equivalent to the results in Valente et al. \cite{cp21}).
This highlights how the dissipative history underlies the dynamic transition undergone by the $\Lambda$ system at zero temperature, both in this semiclassical and in the fully-quantum \cite{cp21} regimes.

At $T \approx 0$ (so that $n_{eb},n_{ea} \ll 1$) and $\Omega > 0$, we find a small (but finite) stationary power consumption 
\begin{eqnarray}
P(\infty) &=& - J(\infty) \approx  E_{ea} \kappa  n_{eb}  \\ 
	&-& \left(3 \kappa E_{ea} + \frac{\kappa^3 E_{ea}}{2 \Omega^2} \right) n_{eb}^2 - \frac{\kappa^3 E_{ea} n_{ea}n_{eb}}{2 \Omega^2},\nonumber
\end{eqnarray}
thus slightly breaking detailed balance (in a way that does not depend on $\Omega$ to first order in the thermal mean number of excitations).
The full expression for the power, at $T\geq0$ and $\Omega > 0$, is given by
\begin{eqnarray}
	P(\infty) = 4 \kappa \Omega^2n_{eb} E_{ea} / \{4 \Omega^2 + 2(\kappa^2 + 6 \Omega^2) n_{eb} \\
	+ \kappa^2 n_{eb}^2 + 2 \kappa^2 n_{ea}^2 (1 + 3 n_{eb}) \nonumber \\ 
	+ \kappa^2 n_{ea} (2 +3 n_{eb}(3 + n_{eb})) \}. \nonumber
\end{eqnarray}

In the numerical calculations, we have set $E_{eb} = 0.99 E_{ea}$.

\section*{$V$ system}
We consider $i = g, b, a$, so that $E_g \ll E_b < E_a$.
The coupling strengths are such that $\Omega_{ga} = \Omega$ and $\Omega_{ba} = \Omega_{gb} = 0$.
The decay rates are $\kappa_{ag} = \kappa_{bg} = \kappa$, and $\kappa_{ab} = 0$.
Similarly, the temperatures are $T_{ag} = T_{bg} = T$.
The steady-state of the $V$ system is uniquely defined for any set of parameters.
At $T = 0$, the system maintains a nonequilibrium coherence 
$\rho_{ag} = -i 2\kappa \Omega/(\kappa^2+8\Omega^2)$ and a nonequilibrium population 
$\rho_{aa} = 4\Omega^2/(\kappa^2 + 8\Omega^2)$, implying that
$P = E_{ag} 4\kappa \Omega^2/(\kappa^2 + 8\Omega^2) = - J_{ag}$,
and $J_{bg} = 0$.
In the numerical calculations, we have set $E_{bg} = 0.99 E_{ag}$.

\section*{$\Diamond$ system}
We consider $i = g, a, b, e$, so that $E_g \ll E_a < E_b \ll E_e$.
The coupling strenghts are such that $\Omega_{gb} = \Omega_{be} = \Omega_{ab} = \Omega_{ge} = 0$.
For the energy-seeking mode, $\Omega_{ga} > 0$ and $\Omega_{ae} = 0$.
For the energy-avoiding mode, $\Omega_{ga} = 0$ and $\Omega_{ae} > 0$.
The decay rates are $\kappa_{ea} = \kappa_{ag} = \kappa_{eb} = \kappa_{bg} = \kappa$, and $\kappa_{ba} = \kappa_{eg} = 0$.
In the numerical calculations, we have set $E_{ag} = 0.9 E_{ea}$, $E_{bg} = 1.1 E_{ea}$, and $E_{eb} = 0.8 E_{ea}$.

In the case of a thermal equilibrium environment, we consider $T_{ea} = T_{ag} = T_{eb} = T_{bg} = T$.
At $T = 0$, the stationary power and the dissipation rates exactly coincide with those in the $\Lambda$-system (for the energy-avoiding mode) and the $V$-system (for the energy-seeking mode).

In the case of thermal gradients, we consider 
$T_L \equiv T_{ea} = T_{ag}$ and $T_R \equiv T_{eb} = T_{bg}$, as well as 
$J_L \equiv J_{ea}+J_{ag}$ and $J_R \equiv J_{eb}+J_{bg}$.

\section{Data availability}
Data sharing not applicable to this article as no datasets were generated or analysed during the current study.

\section{Code availability}
The code used to produce the figures in this article is available from the corresponding author upon request.

\section{Author contributions}
T. W., F. B. and D. V. conceived the main ideas.
T. W. and M. M. performed the analytical and the numerical calculations. 
D. V. wrote the first version of the manuscript.
T. W. prepared the figures.
All authors discussed the results and the final version of the manuscript.
\section{Competing interests}
The authors declare no competing interests.

\begin{acknowledgements}
This work was supported by the Serrapilheira Institute (Grant No. Serra-1912-32056) and by Instituto Nacional de Ci\^encia e Tecnologia de Informa\c c\~ao Qu\^antica (CNPq INCT-IQ 465469/2014-0), Brazil.
M. M. was supported by FAPEMAT.
\end{acknowledgements}

\end{document}